# Design an Advance Computer-Aided
# Tool for Image Authentication and Classification

[1,2]Rozita Teymourzadeh, [1]Amirize Alpha Laadi,
[2]Yazan Samir Algnabi, [2]MD Shabul Islam and [2]Masuri Othman
[1]Department of Electrical and Electronic,
Faculty of Engineering, Technology and Built Environment UCSI University,
Jalan Choo Lip Kung, Taman Taynton View, 56000 Cheras, Kuala Lumpur, Malaysia
[2]Department of VLSI Design, Institute of Micro Engineering and Nanoelectronics IMEN,
Universiti Kebangsaan Malaysia, 43600 Bangi, Selangor, Malaysia

**Abstract: Problem statement:** Over the years, advancements in the fields of digital image processing and artificial intelligence have been applied in solving many real-life problems. This could be seen in facial image recognition for security systems, identity registrations. Hence a bottleneck of identity registration is image processing. **Approach:** These are carried out in form of image preprocessing, image region extraction by cropping, feature extraction using Principal Component Analysis (PCA) and image compression using Discrete Cosine Transform (DCT). Other processing include filtering and histogram equalization using contrast stretching is performed while enhancing the image as part of the analytical tool. Hence, this research work presents a universal integration image forgery detection analysis tool with image facial recognition using Black Propagation Neural Network (BPNN) processor. The proposed designed tool is a multi-function smart tool with the novel architecture of programmable error goal and light intensity. Furthermore, its advance dual database increases the efficiency for high performance application. **Results:** With the fact that, the facial image recognition will always, give a matching output or closest possible output image for every input image irrespective of the authenticity, the universal smart GUI tool is proposed and designed to perform image forgery detection with the high accuracy of ±2% error rate. **Conclusion:** Meanwhile, a novel structure that provides efficient automatic image forgery detection for all input test images for the BPNN recognition is presented. Hence, an input image will be authenticated before being fed into the recognition tool.

## INTRODUCTION

Facial image recognition is a bottleneck of image processing that shows a lot of interrest in past few years (Kumar *et al.*, 2007). Move towards to 2-D face recognition leads to have template-based approach and geometric-based approach (Bolme *et al.*, 2003). further complex features are composed of several of the feature examples. This type of representation leads to an abstraction from the image pixels.

Sirovich and Kirby (1987) proposed Eigen picture as economically representation of image in a best coordinate system. It was observed that an acceptable picture of a face could be reconstructed from the specification of gray levels at 214 pixel locations. However, they showed that in actual construction, only 40 numbers with admixtures of Eigen pictures can be characterized a face.

Wu and Haung (1990) took advantage of 24 measurements in fiducially signs of profile double facial image. The proposed system has been reduced to recognizing performance when more image in the database because there are not enough to differentiate features that identify a user.

Then, Turk and Penland (1991) introduced the Eigen faces method that achieved a remarkable acknowledgement speed, but the speed drastically reduced, when against resizing. Wiskott et al. (1997) used magnitude information and phase information that is received from Gabor wavelet transformation of the face image. For identifying a person, this information is compared with all of the information in database

---

**Corresponding Author:** Rozita Teymourzadeh, Department of Electrical and Electronic, Faculty of Engineering,
Technology and Built Environment, UCSI University, 56000, Cheras Kuala Lumpur Malaysia

searching for the one who has the same transformation as test image.

Gallagher (2005) analyzed a set of images and proposed that image forgery most often includes some form of geometric transformation. These processes were based on a re-sampling and interpolation step.

Later on, Ahonen and Hadid (2006) introduced the Local Binary Pattern (LBP). This operator was used to measure the texture information of local area of gray image.

The various research work had been completed (Ming *et al.*, 2008; Mehryar *et al.*, 2010; Jaeyoung and Jun, 2011; Han *et al.*, 2011; Lu *et al.*, 2003) to reduce the efficiency with variations in illumination, image size, size of database, percentage error. In addition, inability to detect altered images for more realistic recognition still is challenging task.

Hence, this study proposes an efficient passive evaluation tool for authenticity of input images before recognition process stage, furthermore, the adjustment of illumination of images using contrast stretching and image histogram equalization techniques will be performed. In this proposed research work, the quality percentage is adjustable in order to achieve high performance (as low as 2% error rate) with a dual database approach. The test image is reduced to a lower 1-D dimensional vector, which represents distinguishing features of the test image.

The 1-D image vectors are fed into the neural network, the Euclidean distance of the 1-D image vector is compared to each test database and the closest match is found and outputted.

**System design fundamental:** The focus of the image authentication stage (Fig. 1) is the presence of the image re-sampling trace and/or interpolation signal. A derivative operator on the variance of the image will detect these traces. Later, the signal is processed through radon transformation that results in a periodic signal embedded or completely imposed on the image spectrum graph (Gallagher, 2005). Hence, interpolated signal and re-sampled signal will be ready to function.

**Interpolated signal:** There are two main stages in geometric transform (Popescu and Farid, 2005; Prasad & Ramakrishnan, 2006). In the first stage, a spatial transformation of the physical re-arrangement of pixels in the image is performed and can be represented by a transformation function T as shown in Equ.1.

$$X' = T_X(X,Y) \quad Y' = T_Y(X,Y) \qquad (1)$$

As earlier stated, the geometric operations that are common with most forgeries are re-scaling, rotation and skewing. Hence, the importance of detection will be traced wit utilization of affine transformations.

The general equations for affine transformations are given by:

$$X' = a_0 + a_1x + a_2y$$
$$Y' = b_0 + b_1x + b_2y \qquad (2)$$

The second stage is the interpolation. Here pixel intensity values of the transformed image are assigned using low pass interpolation filter. According to the sampling theory, if the Nyquist criterion is satisfied, the spectrum $F(w)$ does not overlap in the Fourier domain. The original signal $f(x)$ can be reconstructed perfectly from its samples $f_k$ using the optimal sine interpolation (Hou and Andrews, 1987).

Combining the derivative theorem with the convolution theorem leads to the conclusion that by convolution of $f_k$ with derivative kernel $D_n(w)$, it is possible to reconstruct the nth derivative of the image $f(x)$. The result of interpolated operation of $f_w(x)$ is denoted by Equ. 3;

$$f^w(x) = \sum_{k=-\infty}^{\infty} f_k w\left(\frac{x}{\Delta x} - k\right)$$
$$D^n\{f^w\}(x) = D^n\left\{\sum_{k=-\infty}^{\infty} f_k w\left(\frac{x}{\Delta x} - k\right)\right\} =$$
$$\sum_{k=-\infty}^{\infty} f_k D^n\{w\}\left(\frac{x}{\Delta x} - k\right) \qquad (3)$$

By assuming the constant variance random process, then the variance of $D^n\{f^w\}$ which is var $\{D_n\{f_w\}(x)\}$ as a function of $x$ is given by:

$$var\{D^n\{f^w\}(x)\} = R_{D^n\{f^w\}}(x,x) =$$
$$\sigma^2 \sum_{k=-\infty}^{\infty} D^n\{w\}\left(\frac{x}{\Delta x} - k\right)^2 \qquad (4)$$

Similarly, the covariance is represented as Equ. 5:

$$R_{D^n\{f^w\}}(x, x+\xi) = \sigma^2 \sum_{k=-\infty}^{\infty} D^n\{w\}\left(\frac{x}{\Delta x} - k_1\right)$$
$$xD^n\{w\}\left(\frac{x+\xi}{\Delta x} - k_2\right) \qquad (5)$$

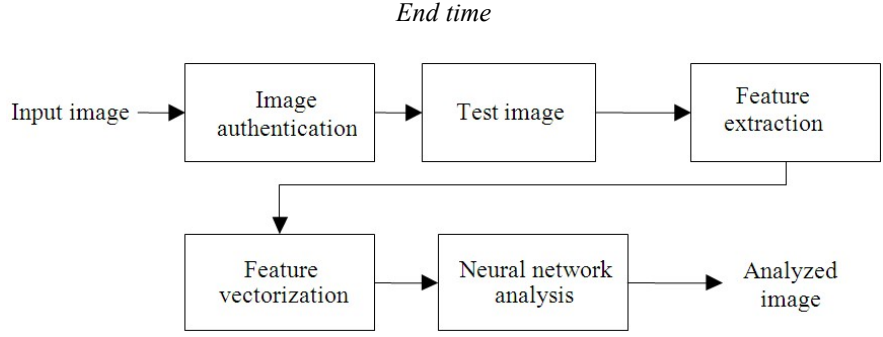

*End time*

Fig. 1: Proposed process structure

Now, by assuming that $\vartheta$ is an integer, it can be noticed that:

$$var\{D^n\{f^w\}(x)\} = var\{D^n\{f^w\}(x + \vartheta \Delta)\} \quad (6)$$

Thus, $var\{D_n\{f_w\}(x)\}$ is periodic over $x$ with period $\Delta x$ that $\Delta x$ is the sampling step.

It is verified the periodicity by the following:

$$var\{D^n\{f^w\}(x + \vartheta \Delta)\}$$
$$= \sigma^2 \sum_{k=-\infty}^{\infty} D^n\{w\} \left(\frac{x + \vartheta \Delta x}{\Delta x} - k\right)^2$$
$$= \sigma^2 \sum_{k=-\infty}^{\infty} D^n\{w\} \left(\frac{x}{\Delta x} - (k - \vartheta)\right)^2 = var\{D^n\{f^w\}(x)\}$$
$$\quad (7)$$

**Re-sampled signal:** From Eq. 4, it is clear that different interpolators will change the structure of the original signal in different ways. The resulting periodic variance function computed using Eq. 4 for the nearest-neighbor interpolation. Hence, signals interpolated by this interpolator are easily recognized by applying a derivative operator to them (Hou and Andrews, 1987). It is continuous but its first derivative is discontinuous. Cubic interpolation is a very frequently used interpolation technique and has been widely studied. It uses a third order interpolation polynomial as the kernel.

**Radon transformation:** Radon transformation is applied to find traces of affine transformation. The Radon transformation computes projections of magnitude of $D_n\{b(x, y)\}$ along specified direction determination by an angle $\theta$.

The projection is a line integral in a certain direction. This line integral is expressed as:

$$\rho D^n\{b\}(x, y) = \int_L |D^n\{b(x,y)\}| \, dl \quad (8)$$

By assuminy that:

$$\begin{bmatrix} x' \\ y' \end{bmatrix} = \begin{bmatrix} cos\Theta & sin\Theta \\ -sin\Theta & cos\Theta \end{bmatrix} \begin{bmatrix} x \\ y \end{bmatrix} \quad (9)$$

It is possible to represent the Radon transform in the following way:

$$\rho\theta(x') = \int_{-\infty}^{\infty} D^n\{b(x, y)\} \cdot (x' cos\Theta - y' sin\Theta, x' sin\Theta + y' cos\Theta) dy'$$
$$\quad (10)$$

The Radon transformation is computed at angles $\theta$ from 0-179 degrees, in 1 degree increments. Hence, the output of this is 180 1-D vectors, $\delta\theta$ ($\theta$ is the orientation of the $X'$ axis counterclockwise from the $x$-axis). The corresponding auto covariance sequences of $\delta\theta$ contain a specific strong periodicity, if the investigated signal has been re-sampled. The auto covariance sequences of $\rho\theta$ is computed as Eq.11:

$$R_{\rho\theta}(k) = \sum_i (\rho\theta\,(i+k) - \overline{\rho\theta})(\rho\theta\,(i) - \overline{\rho\theta}) \quad (11)$$

The project target is to determine the image being investigated has undergone affine transformation. Hence, we focus only on the strongest periodic patterns present in the auto covariance $R\delta\theta$.

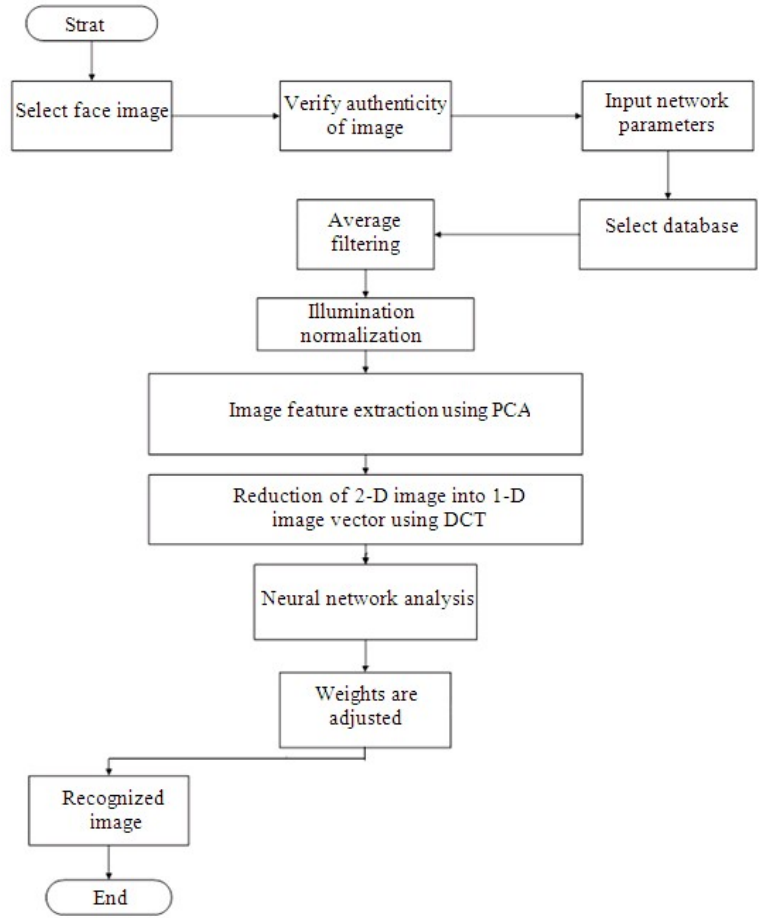

Fig. 2: Proposed system performance flowchart

**Image recognition:** The difficulty of recognizing faces is a classification problem. The proposed technique uses artificial neurons in order to train the classifier. It also involves a dimensionality reduction preprocessing of the facial images. In recent times, BPNN system (Vinay *et al.*, 2007) implementation make use of fractal encoding method, the fractal codes were presented as input to the BPNN for identification purposes. However, in order to complete image authentication and recognition in the research work, four stages are introduced.

**Proposed system structure:** The proposed research work deals with an integration of four phases. The initial phase deals with image verification (originality authentication) with neural network pre-processing training and testing phases. Figure 2 shows the flowchart performance of this research work that will be discussed in four stages.

**Image verification (Phase 1):** The image verification phase is performed using detecting traces of image tampering such as re-sampling or interpolation or both. In this phase, a test/input process involves derivative operators and radon transformation. This process produces a periodic pattern in the image spectrum if the test image is forged or it produces simple impulse signals if the test image is still original.

Meanwhile, It is observed that both image spectra forms could be overlapped that produces a form of image tampering. However, the proposed system specification allows selecting only original test or input images.

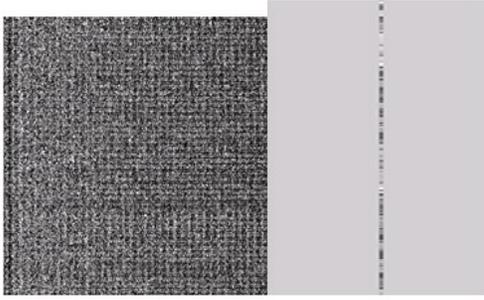

Fig. 3: Image vectorization

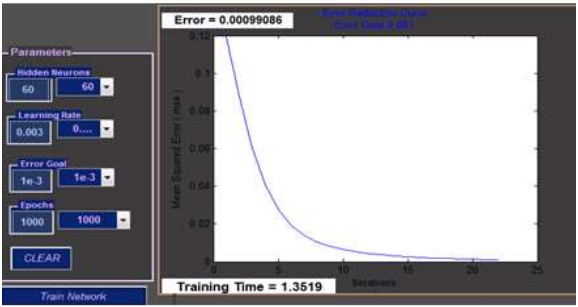

Fig. 4: Neural network training curve

**Image-preprocessing (Phase 2):** The image preprocessing as second phase is ready to function. In the pre-processing phase, time effective preprocessing is performed in order to make image data best fit for neural network input. Average filtering is applied and contrast of the image is enhanced through histogram equalization process.

Then the image size is reduced in order to make it light but efficient and best fit for neural processing phase. Before moving images into the neural processing phase, all images data (test image and training database images) undergo a process of vectorization that is conversion of 2-D images into 1-D vectors. This conversion is due to neural network requires 1-D vectors for processing and conversion is performed using PCA (Esbensen and Geladi, 1987) and 2-D discrete cosine transform. Figure 3 shows the vectorization of one original image.

**Neural network training (Phase 3):** The third phase is the neural network training. The structure of the proposed neural network is based upon multi-layers that are input, hidden and output layer. The number of neurons in the hidden layer is determined by experiences and guesswork considering optimal performance.

As a final point, production level contains nerve cells that are the number of objects is being considered. The algorithm for effective use of a neural networks and reduces the slope of errors through changing weight and offset continued with impetus. The neural network is trained upon some set of images and tested upon different set of images. In the proposed method, neural network uses BPNN algorithm for error computation and new weight calculation for each neuron link. It returns the output of each level, extract the mean square error (MSE) and spread it back if it is not close to the target. The response of the neural network is reliant upon weights, biases and transfer functions. The transfer functions make use of in the feed forward BPNN in intermediate, input and output layer. Fig. 4 shows the training graph of neural network while the system will find the original image.

**Testing stage (Phase 4):** The fourth phase is the testing of the neural network. Images for testing are applied to the trained neural network along with the already trained database images for calculating the percentage accuracy and error.

In testing phase, just like in the training phase, incoming images undergo all the pre-processing stage and are made available to the network for evaluation. This test images (extracted face image) is then processed using the neural network analytic tool. After a number of iterations by the network through each image in the database, the error reduces based on the gradient descent-learning rule, the set error goal is reached and a matching image to test image is found as shown in the results session.

## RESULTS AND DISCUSSION

The two database containing training images is made up of ten images numbered (1-10) each, with the same size, format and dimension. The test images are sampled and cropped with the different position. Figure 5 and 6 show the image spectrum respectively.

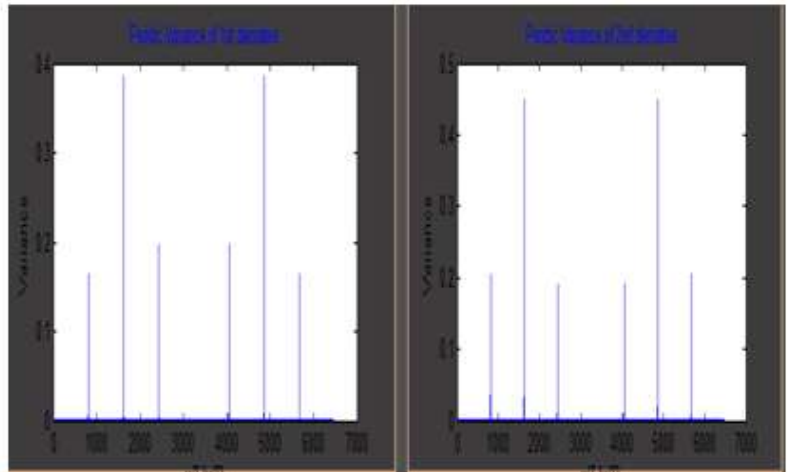

Fig. 5:  Original image spectrum

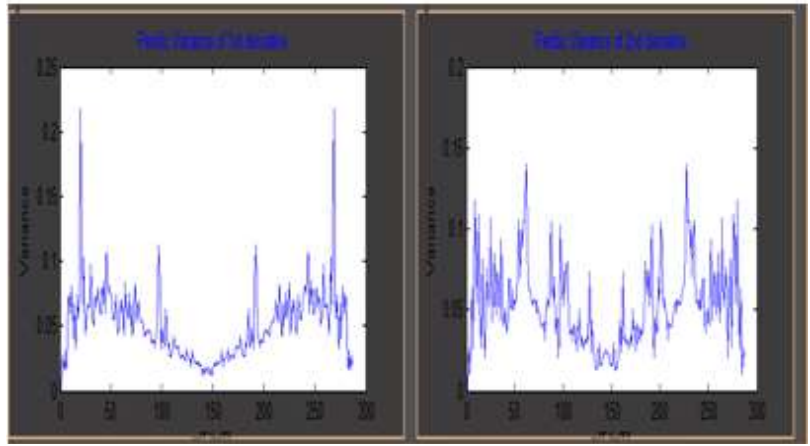

Fig. 6:  Forged image spectrum

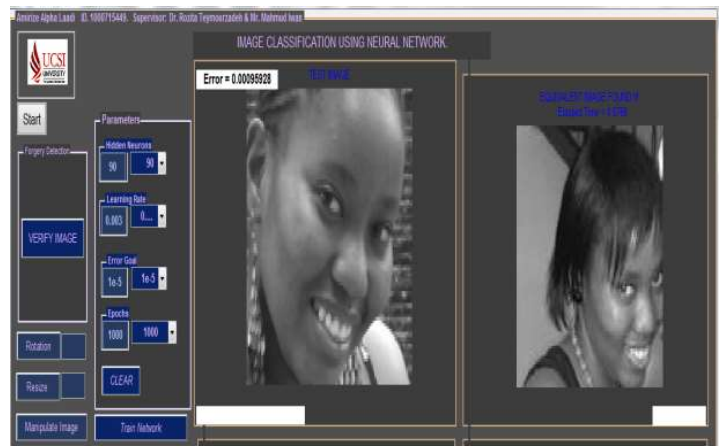

Fig. 7: Image recognition

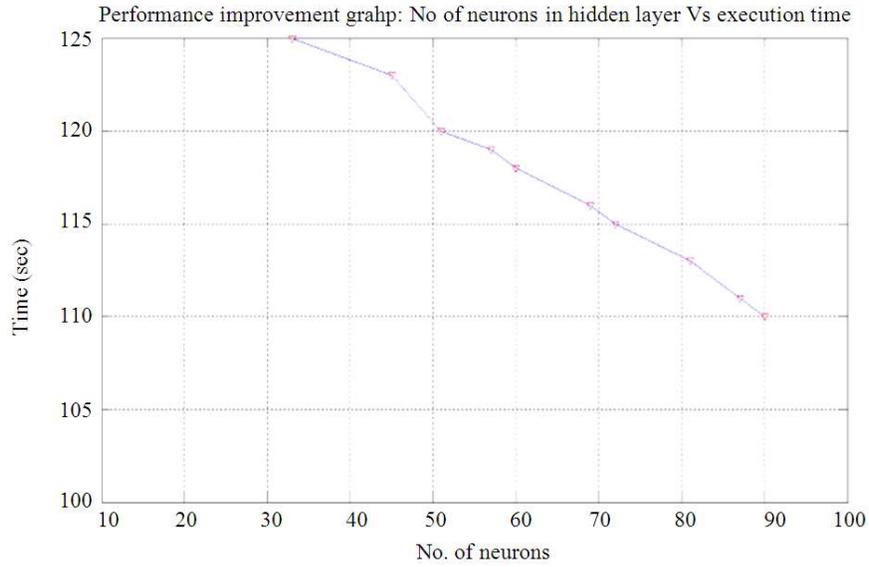

Fig. 8: Neurons in hidden layer vs. execution time

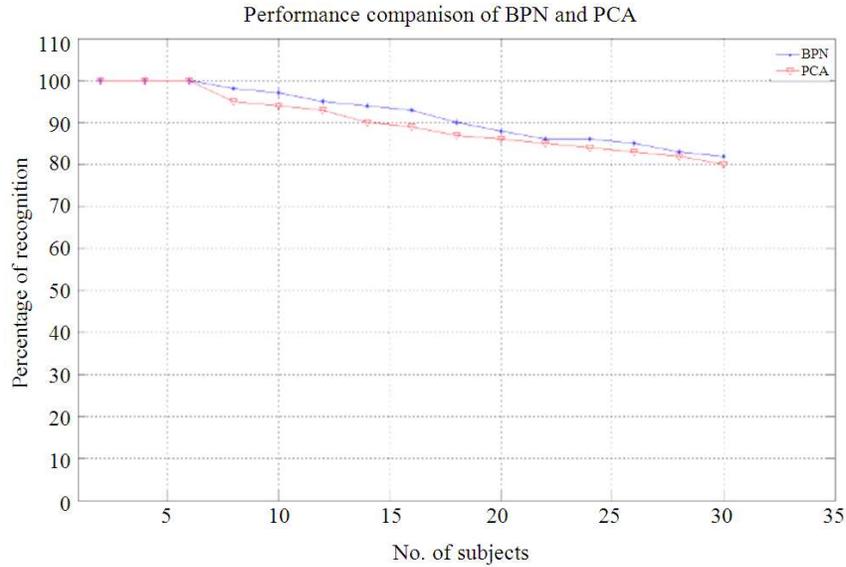

Fig. 9: Recognition percentage vs. number of subjects

These spectra are resulted from original and forged images respectively. The difference in spectrum is seen as a kind of sinusoidal periodic pattern, which replaces or overlaps the original impulse signals or spikes of an original image. These are the result of obtaining the first and second derivative of the auto covariance of the images. The auto covariance is obtained using the radon transformation through 0-179 degrees. Hence, it is a proved for efficiency of proposed system to detect the authenticity of images.

The major integral part of the proposed analytical tool is the capability to extract a face from a full image and run it through any given database for a matching or equivalent face image. Figure 7 illustrates the recognition of input image while different position is applied.

The high-speed performance is determined by the proper selection of the number of hidden neurons. The larger amount of hidden neurons results the faster network converges. Figure 8 shows the time consuming required when the hidden layers in neural network processor are increased.

Figure 9 shows the error percentages of proposed analytical tools when the number of imaged are increased. As shown in Fig. 9, it is found that with increasing the number of subjects, performance error that is percentage of recognition with utilizing PCA and BPNN approach will decrease slightly.

## MATERIALS AND METHODS

The analytical image authentication tool designed based on graphical interfacing computer-aided system. The design process was to create software model of efficient tool using MATLAB-GUI softwar. The system was designed simulated for image authentication and classification. The design was emulated the following characteristics; neureal network processing, local binary pattern and finaly analytical tools smart system.

## CONCLUSION

In this research work, an integration of image forgery detection with an image facial recognition analytical tool using back propagation neural network was proposed. In this project the test image is authenticated before being fed into the recognition tool. Therefore, the result of the first image verification of this project confirms the originality or alteration of an image before it is feed into the rest of the algorithm for recognition purpose. It also introduces contrast stretching of histograms, dual database and adjustable percentage error as a functionality of the proposed research work. The BPNN for face image recognition is only highly accurate with small number of subjects and requires a lowering of the image resolution for a whole recognition task as a way of reducing the computational complexity. The project was designed and investigated and it was found that the resolution improved by ±2% error rate when number of database images applied is less than 20 images.

## REFERENCES

Ahonen, T. and A. Hadid, 2006. Face recognition with local binary patterns: Application on Face Recognition, Proc. IEEE Transactions on Pattern Analysis and Machine Intelligence IEEE Press, 28: 2037-2041. DOI: 10.1109/TPAMI.2006.244

Bolme, D., R. Beveridge, M. Teixeira and B. Draper, 2003. The CSU Face Identification Evaluation System: Its Purpose, Features and Structure. International Conference on Vision Systems, pp: 304-311. Graz, Austria, April 1-3. Published by Springer-Verlag. DOI: 10.1.1.89.1918

Esbensen K, Geladi P. 1987. Chemometrics and Intelligent Laboratory Systems. Proceedings of the Multivariate Statistical Workshop for Geologists and Geochemists, Volume 2, Issues 1–3, pp. 37–52. DOI: 10.1016/0169-7439(87)80084-9

Gallagher, A.C., 2005. Detection of linear and cubic interpolation in JPEG compressed images. In Proceeding of the IEEE Computer. Soc 2nd Canadian Conf. Computer Robot Vision, Washington, DC, pp: 65-72. DOI: 10.1109/CRV.2005.33

Han, H., J. Jeong and E. Arai, 2011. Virtual out of focus with single image to enhance 3D perception. IEEE Conference on 3DTV The True Vision-Capture, Transmission and Display of 3D Video, pp: 1-4 DOI: 10.1109/3DTV.2011.5877188

Hou, H. and H. Andrews, 1978. Cubic splines for image interpolation and digital filtering. IEEE Trans. Acoust., Speech Signal Process., 26: 508-517. DOI: 10.1109/TASSP.1978.1163154

Jaeyoung, K. and H. Jun, 2011. Implementation of image processing and augmented reality programs for smart mobile device. IEEE Conference on Strategic Technology (IFOST). 2: 1070-1073. DOI: 10.1109/IFOST.2011.6021205

Kumar, V.B., B.S. Shreyas and G.C.N.S. Murthy, 2007. A Back Propagation Based Face Recognition Model, Using 2D Symmetric Gabor Features. IEEE Proceeding of The Signal Processing, Communications And Networking, pp: 433-437. DOI: 10.1109/ICSCN.2007.350776

Lu, J., K.N. Plataniotis and A.N. Venetsanopoulos, 2003. Regularized Discriminate Analysis For the Small Sample Size Problem in Face recognition, Science Direct Pattern Recognition Letters. Sci. Direct 24: 3079-3087. DOI: 10.1016/S0167-8655(03)00167-3

Mehryar, S., K. Martin, K.N. Plataniotis and S. Stergiopoulos, 2010. Automatic landmark detection for 3D face image processing. IEEE Conference on Evolutionary Computation (CEC), pp: 1-7. DOI: 10.1109/CEC.2010.5586520

Ming, H., Q. Zhang and Z. Wang, 2008. Application of Rough Sets to Image Pre-Processing for Face Detection. IEEE Proc. Of the information and automation, pp: 545-548. DOI: 10.1109/ICINFA.2008.4608060

Popescu A.C. and H. Farid, 2005. Exposing digital forgeries by detecting traces of re-sampling. IEEE Trans. Signal Process., 53: 758-767. DOI: 10.1109/ICOSP.2006.345714

Prasad, S. and K.R. Ramakrishnan, 2006. On re-sampling detection and its application to image tampering. in cedinging of the IEEE Interenational Conference Multimedia Expo., Toronto, ON, Canada, pp: 1325-1328. DOI: 10.1109/ICIEA.2009 5138406

Sirovich, L. and M. Kirby, 1987. A low dimensional procedure for the characterization of human faces. J. Optical Society Am., 4:519-524. DOI: 10.1364/JOSAA.4.000519

Turk, M.A. and A.P. Pentland, 1991. Face recognition using eigenfaces. IEEE Conference on Computer Vision and Pattern Recognition, pp: 586-591. DOI: 10.1109/CVPR.1991.139758

Wiskott, L., J.M. Fellous, N. Kuiger and C. von der Malsburg, 1997. Face recognition by elastic bunch graph matching. IEEE Trans. On pattern Analysis and machine intelligence, 19: 775-779. DOI: 10.1109/34.598235

Wu, C.J. and J.S. Huang, 1990. Human face profile recognition by computer. J. Sci. Direct Pattern Recognition, 23: 255-259. DOI: 10.1016/0031-3203(90)90013-B